\documentclass[twocolumn,pra,superscriptaddress,showpacs]{revtex4-2}

\usepackage{amsmath,amssymb,amsfonts}
\usepackage{graphicx}
\usepackage{subcaption}
\usepackage{hyperref}
\usepackage{xcolor}
\usepackage{physics}
\usepackage{siunitx}
\usepackage{bm}
\usepackage{float}

\hypersetup{
    colorlinks=true,
    linkcolor=blue,
    citecolor=blue,
    urlcolor=blue
}

\newcommand{\Vdd}{V_{\mathrm{dd}}}
\newcommand{\cm}{\mathrm{cm}^{-1}}

\begin{document}

\title{Noise-Robust Ultrafast Entanglement Generation in Rydberg Atoms via Quantum Optimal Control}

\author{Tanveer Ahmad}
\email{tanveerahmad@qq.com}
\affiliation{Hunan Key Laboratory of Super-Microstructure and Ultrafast Process, 
School of Physics, Central South University, Changsha 410083, China}

\date{\today}

\begin{abstract}
We present a comprehensive theoretical analysis of ultrafast entanglement generation between two Rydberg-blockaded atoms, explicitly accounting for realistic laser noise. Using femtosecond Gaussian pulses as a baseline, we systematically evaluate Bell-state fidelity sensitivity to amplitude and phase noise across white, pink~($1/f$), and Ornstein--Uhlenbeck spectra using Monte Carlo ensemble simulations. Our results show that amplitude noise is well tolerated, with fidelities above 90\% even at 30\% noise levels, while phase noise is the primary limiting factor, causing fidelity to drop rapidly beyond $\sim$1\% noise amplitude. The spectral structure of the noise is also important: pink noise consistently causes less fidelity loss than white noise of the same amplitude. By applying quantum optimal control theory (QOCT) with the D-MORPH algorithm under multiple equality constraints, we obtain a double-pulse structure with a spectral notch that achieves approximately 99\% fidelity in the noise-free case and maintains high fidelity under moderate amplitude noise. A breakdown threshold near $\sim$1\% amplitude noise is identified, beyond which even optimized pulses cannot sustain coherent control. These results offer practical benchmarks for the development of ultrafast neutral-atom quantum processors operating in the femtosecond regime.
\end{abstract}


\maketitle

\section{Introduction}
\label{sec:intro}

Neutral atoms in optical tweezer arrays are now a leading platform for quantum information processing. Recent experiments have demonstrated two-qubit gate fidelities above 99.5\%~\cite{Evered2023Nature,Scholl2023Nature} and fault-tolerant architectures with up to 448 atoms~\cite{Bluvstein2024Nature}. These results depend on strong, long-range dipole-dipole interactions between atoms in highly excited Rydberg states, which enable deterministic entanglement via the Rydberg blockade mechanism~\cite{Jaksch2000,Saffman2010}. While current systems operate on microsecond timescales, there is increasing interest in achieving femtosecond-scale operation to reduce decoherence and approach the quantum speed limit.

Ultrafast quantum control faces a key challenge: faster operations require broader spectral bandwidths and higher peak intensities, which increase sensitivity to laser noise. Amplitude and phase fluctuations are common in experiments~\cite{Jiang2023PRA,Ball2022} and are a major error source for neutral-atom qubits. Understanding their impact on ultrafast entanglement and developing effective mitigation strategies is essential to realizing femtosecond-scale quantum gates.

Most theoretical studies of Rydberg entanglement address ideal, noise-free conditions~\cite{Guo2019PRA, Shu2016PRA} or specific noise models in longer-timescale gates~\cite{Levine2018,Levine2019}. A systematic analysis of noise effects in ultrafast Rydberg dynamics, particularly regarding noise spectral structure, remains lacking. While quantum optimal control theory (QOCT) has been used to design robust pulses in other quantum systems~\cite{Koch2022,Muller2022,Aroch2024Quantum}, its application to noise-robust ultrafast Rydberg entanglement is not yet well developed.

In this work, we address these gaps by combining systematic noise analysis with optimal pulse design. We perform a systematic Monte Carlo analysis of ultrafast Rydberg entanglement under amplitude and phase noise with white, pink~($1/f$), and colored (Ornstein--Uhlenbeck) spectra, identifying amplitude noise tolerance near 30\%, phase noise sensitivity thresholds around 1\%, and key spectral effects. Using quantum optimal control with the D-MORPH algorithm under multiple equality constraints, we generate optimized pulse structures that maintain high fidelity in the presence of moderate noise. We also establish breakdown thresholds and provide practical benchmarks relevant to experimental implementation of ultrafast Rydberg gates.

The paper is organized as follows. Section~\ref{sec:theory} presents the theoretical framework including the atom Hamiltonian, noise models, and optimal control formalism. Section~\ref{sec:noise_analysis} details our systematic noise sensitivity analysis. Section~\ref{sec:qoct} presents the optimized pulse design and its performance under noise. Section~\ref{sec:discussion} discusses experimental implications and connections to recent advances. Section~\ref{sec:conclusion} concludes.

\section{Theoretical Framework}
\label{sec:theory}

\subsection{Two-Atom Model and Effective Three-Level System}
\label{subsec:hamiltonian}

We consider a system of two identical two-level atoms in controlled arrays of optical dipole traps, following the model of Guo~\textit{et al.}~\cite{Guo2019PRA}. Each atom consists of two levels $|0\rangle$ and $|1\rangle$ with transition frequency~$\omega_0$ and dipole moment~$\mu$, corresponding to the $5S_{1/2}$ and $5P_{1/2}$ states of $^{87}$Rb. The atoms interact through Ising coupling, and the field-free Hamiltonian reads ($\hbar = 1$)
\begin{equation}
H_0 = \frac{\omega_0}{2}\sum_{i=1}^{2} S_i^z + \Vdd \sum_{i\neq j} S_i^+ S_j^-,
\label{eq:H0}
\end{equation}
where $S_i^+$, $S_i^-$, and $S_i^z$ are the raising, lowering, and energy difference operators of the $i$th atom, and $\Vdd$ is the dipole-dipole interaction strength.

\begin{figure}[htbp]
\centering
\includegraphics[width=\columnwidth]{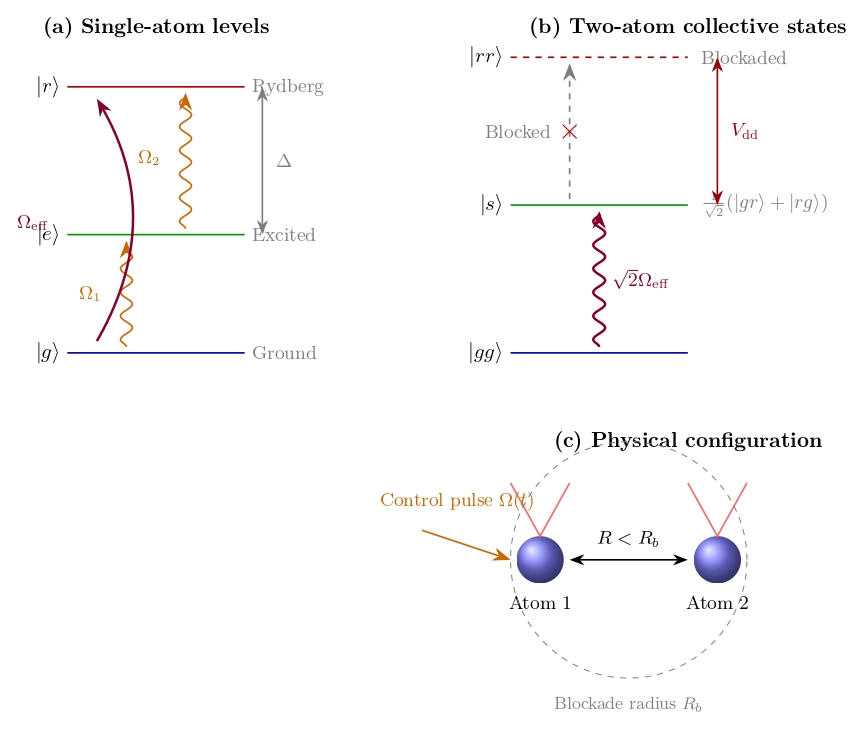}
\caption{Rydberg blockade mechanism for two-atom entanglement. (a)~Single-atom energy levels with two-photon coupling via Rabi frequencies $\Omega_1$ and $\Omega_2$. (b)~Two-atom collective states showing the symmetric entangled state $|s\rangle = (|01\rangle + |10\rangle)/\sqrt{2}$ and the blockaded doubly-excited state shifted by dipole-dipole interaction~$\Vdd$. (c)~Physical configuration of two atoms separated by distance $R < R_b$, where $R_b$ is the blockade radius.}
\label{fig:level_scheme}
\end{figure}

In the presence of dipole-dipole coupling, the system eigenstates are represented in the Dicke basis: $|g\rangle = |00\rangle$ (ground), $|a\rangle = (|01\rangle - |10\rangle)/\sqrt{2}$ (antisymmetric), $|s\rangle = (|01\rangle + |10\rangle)/\sqrt{2}$ (symmetric), and $|e\rangle = |11\rangle$ (doubly excited), with eigenvalues $E_g = -\omega_0$, $E_a = -\Vdd$, $E_s = +\Vdd$, and $E_e = +\omega_0$, respectively. The antisymmetric state $|a\rangle$ decouples from the laser-driven dynamics~\cite{Almutairi2011,Stefanatos2019}, reducing the system to a three-level ladder with Hamiltonian
\begin{equation}
H_d(t) = \sum_{p=g,s,e} |p\rangle E_p \langle p| - \mu_d \mathcal{E}(t) \!\left(\sum_{\substack{p,q\\=g,s}} |p\rangle\langle q| + \sum_{\substack{p,q\\=s,e}} |p\rangle\langle q|\right),
\label{eq:Hd}
\end{equation}
where $\mu_d = \sqrt{2}\mu$ is the effective transition dipole moment and $\mathcal{E}(t)$ is the time-dependent control field.

For our system parameters, we use $\omega_0 = 12\,578.95~\cm$, $\mu = 7.61$~Debye, and $\Vdd = 12.35~\cm$ (corresponding to interatomic separation $d = 100$~a.u. and $\alpha = \pi/2$).

\subsection{Entanglement Generation and Pulse Area Theorem}
\label{subsec:blockade}

The goal is to drive the system from $|g\rangle$ at $t = t_0$ to the maximally entangled Bell state
\begin{equation}
|s\rangle = \frac{1}{\sqrt{2}}\left(|01\rangle + |10\rangle\right)
\label{eq:bell_state}
\end{equation}
at $t = t_f$, with fidelity $F = |\langle s|\psi(t_f)\rangle|^2$. By expanding the time-dependent unitary operator to the first leading term of the Magnus expansion~\cite{Blanes2009,Shchedrin2015}, pulse area integrals $\theta_{sg}(t) = \mu_d \int \mathcal{E}(t') e^{i\omega_{sg}t'} dt'$ and $\theta_{es}(t) = \mu_d \int \mathcal{E}(t') e^{i\omega_{es}t'} dt'$ determine the population dynamics~\cite{Guo2019PRA}. Perfect transfer to $|s\rangle$ requires $\theta_{sg}(t_f) = \pi/2$ and $\theta_{es}(t_f) = 0$; the first condition sets the pulse area for resonant excitation, and the second suppresses leakage to $|e\rangle$.

When the laser bandwidth $\Delta\omega \ll \Vdd$, narrow-band pulses satisfy both conditions and the system behaves as an effective two-level model ($|g\rangle \leftrightarrow |s\rangle$), exhibiting Rabi oscillations with fidelity $F > 0.9999$. For ultrafast pulses with durations $T \sim 100$--$400$~fs, the bandwidth $\Delta\omega \sim 1/T$ exceeds~$\Vdd$, breaking the two-level description and allowing population leakage to $|e\rangle$. This motivates the use of quantum optimal control to find constrained pulse shapes that restore high fidelity in the broadband regime.

\subsection{Laser Noise Models}
\label{subsec:noise}

Real laser systems exhibit fluctuations in both amplitude and phase~\cite{Jiang2023PRA,Ball2022}. We model noisy pulses as
\begin{equation}
\mathcal{E}(t) = \mathcal{E}_0(t)\left[1 + \epsilon_A \eta_A(t)\right] e^{i\epsilon_\phi \eta_\phi(t)},
\label{eq:noisy_field}
\end{equation}
where $\mathcal{E}_0(t)$ is the ideal pulse envelope, $\epsilon_A$ and $\epsilon_\phi$ are the noise amplitudes, and $\eta_A(t)$, $\eta_\phi(t)$ are stochastic processes with unit variance. Throughout this work, the noise amplitude is denoted $\alpha$ in figures, where $\alpha \equiv \epsilon_A$ for amplitude noise and $\alpha \equiv \epsilon_\phi$ for phase noise.

We consider three noise spectral types. \textit{White noise} has a flat power spectral density $S_{\mathrm{white}}(\omega) = S_0$. \textit{Pink noise}~($1/f$) has scale-invariant fluctuations with $S_{\mathrm{pink}}(\omega) \propto 1/|\omega|^\beta$ ($\beta \approx 1$), ubiquitous in electronic systems. \textit{Ornstein--Uhlenbeck~(OU) noise} is colored noise with finite correlation time~$\tau_c$ and Lorentzian spectrum $S_{\mathrm{OU}}(\omega) = 2\sigma^2\tau_c / (1 + \omega^2\tau_c^2)$. Noise realizations are generated via spectral shaping in the frequency domain and transformed back to the time domain with proper normalization (see Appendix~\ref{app:noise}).

\subsection{Quantum Optimal Control via D-MORPH}
\label{subsec:qoct}

To design pulse shapes that maximize fidelity while satisfying physical constraints, we employ the D-MORPH algorithm developed by Shu, Ho, and Rabitz~\cite{Shu2016PRA}. This method introduces a dummy variable~$s$ along which the control field $\mathcal{E}(s,t)$ is updated from an initial guess $\mathcal{E}(0,t)$ to an optimized field while monotonically increasing the objective and exactly preserving multiple equality constraints $h_m[\mathcal{E}(s,\cdot)] = C_m$ ($m = 1,\ldots,M$).

The constrained update equation is
\begin{equation}
\frac{\partial \mathcal{E}(s,t)}{\partial s} = S(t)\sum_{\ell=0}^{M} [\Lambda^{-1}]_{0\ell}\, c_\ell(s,t),
\label{eq:dmorph}
\end{equation}
where $S(t) \geq 0$ is a smooth envelope function restricting the field variation, $c_0(s,t) = \delta F / \delta\mathcal{E}(s,t)$ is the objective gradient, $c_m(s,t) = \delta h_m / \delta\mathcal{E}(s,t)$ for $m = 1,\ldots,M$ are the constraint gradients, and $\Lambda$ is a Gram matrix with elements $\Lambda_{\ell\ell'} = \int S(t)\, c_\ell(s,t)\, c_{\ell'}(s,t)\, dt$. The objective gradient is computed via the adjoint-state method~\cite{Shu2016PRA}:
\begin{equation}
\frac{\delta F}{\delta\mathcal{E}(s,t)} = -2\,\mathrm{Im}\!\left(\mathrm{Tr}\left\{[|i\rangle\langle i|, O(T)]\,\mu(t)\right\}\right),
\label{eq:gradient}
\end{equation}
with $\mu(t) = U^\dagger(t,0)\,\mu\, U(t,0)$ and $O(T) = U^\dagger(T,0)|f\rangle\langle f|U(T,0)$.

We impose three simultaneous equality constraints on the optimized field: (i)~zero pulse area $h_1 = \int \mathcal{E}(t)\,dt = 0$, removing the dc component for all-optical control; (ii)~constant fluence $h_2 = \int \mathcal{E}^2(t)\,dt = \mathrm{const}$, fixing the total pulse energy; and (iii)~fixed spectral pulse area $h_3 = \mu_d\int \mathcal{E}(t)\cos(\omega_{sg} t)\,dt = \pi/2$, enforcing the resonance condition of the effective two-level transition. These three constraints yield $c_1 = 1$, $c_2 = 2\mathcal{E}(s,t)$, and $c_3 = \mu_d \cos(\omega_{sg}t)$, respectively. The simultaneous enforcement of all three constraints throughout the optimization is a key feature of this approach.

\section{Noise Sensitivity Analysis}
\label{sec:noise_analysis}

\subsection{Simulation Methodology}
\label{subsec:methodology}

We perform Monte Carlo simulations of the time-dependent Schr\"{o}dinger equation under noisy driving fields. For each noise configuration (type, amplitude), we generate 100 independent noise realizations, propagate the three-level wavefunction using matrix exponentiation of the time-dependent Hamiltonian at each time step [i.e., $U(t_{j+1}, t_j) = \exp(-i\,H(t_j)\,\Delta t)$, computed via the \texttt{expm} function], compute the fidelity $F$ for each realization, and extract the mean fidelity $\bar{F}$ and standard deviation $\sigma_F$.

The baseline (noise-free) pulse is a Gaussian
\begin{equation}
\mathcal{E}_0(t) = \frac{\sqrt{\pi/2}}{\mu_d\,\tau} \exp\!\left[-\frac{t^2}{2\tau^2}\right] \cos(\omega_{sg}\, t),
\label{eq:gaussian_pulse}
\end{equation}
designed such that $\theta_{sg}(t_f) = \pi/2$ for any duration~$\tau$. We examine three durations: $\tau = 100$~fs ($\Delta\omega \approx 4.0\,\Vdd$), $\tau = 250$~fs ($\Delta\omega \approx 1.7\,\Vdd$), and $\tau = 400$~fs ($\Delta\omega \approx 1.0\,\Vdd$), with a time grid of $N_t = 10\,000$ steps over $[-4\tau, 4\tau]$.

We compare three model descriptions: \textbf{3LN}---full three-level numerical simulation via matrix exponentiation of Eq.~(\ref{eq:Hd}); \textbf{3LA}---three-level analytical model using the first-order Magnus expansion; and \textbf{2LA}---reduced two-level analytical approximation neglecting $|e\rangle$.

\subsection{Amplitude Noise Tolerance}
\label{subsec:amplitude_noise}

\begin{figure}[htbp]
\centering
\includegraphics[width=\columnwidth]{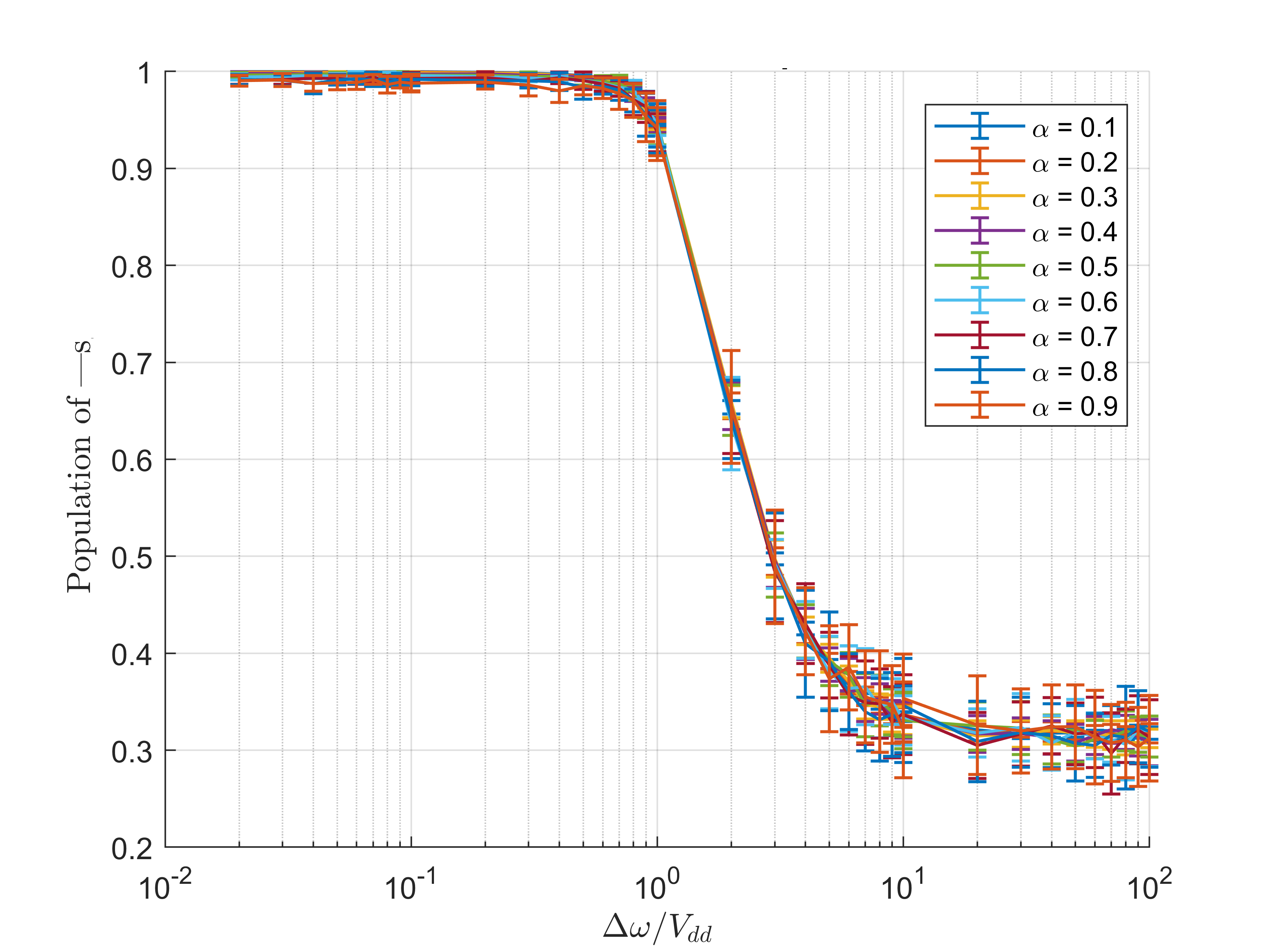}
\caption{Target state population $P_s = |\langle s|\psi(T)\rangle|^2$ versus normalized pulse bandwidth $\Delta\omega/\Vdd$ under white amplitude noise for $\alpha = 0.1$--$0.9$ (where $\alpha \equiv \epsilon_A$). Population remains above 90\% even at 30\% noise in the blockade regime ($\Delta\omega/\Vdd \ll 1$), demonstrating remarkable tolerance to amplitude fluctuations. Error bars show standard deviation over 100 Monte Carlo realizations.}
\label{fig:amplitude_noise}
\end{figure}

Figure~\ref{fig:amplitude_noise} shows the population in the target state $|s\rangle$ as a function of normalized bandwidth $\Delta\omega/\Vdd$ under white amplitude noise for several noise amplitudes. The key finding is the remarkable tolerance to amplitude fluctuations: population remains above 90\% even at $\alpha \equiv \epsilon_A \approx 0.3$ (30\% amplitude fluctuation).

This robustness can be understood through the pulse area theorem~\cite{Allen1975}. For the effective two-level system ($|g\rangle \leftrightarrow |s\rangle$), the final state depends primarily on the integrated pulse area $\Theta = \int \Omega(t)\,dt$. Random amplitude fluctuations tend to average out over the pulse duration, leaving the mean area largely unchanged. The fidelity degradation follows approximately $F \approx F_0(1 - c_A\epsilon_A^2)$, where $c_A$ is a model-dependent coefficient, explaining the gradual, quadratic degradation observed.

The spectral structure of amplitude noise has a surprisingly weak effect: white and pink noise produce nearly identical fidelity curves at matched amplitudes. This occurs because amplitude noise enters multiplicatively, coupling to the instantaneous field strength rather than the accumulated phase.

\subsection{Phase Noise Sensitivity}
\label{subsec:phase_noise}

\begin{figure}[htbp]
\centering
\includegraphics[width=\columnwidth]{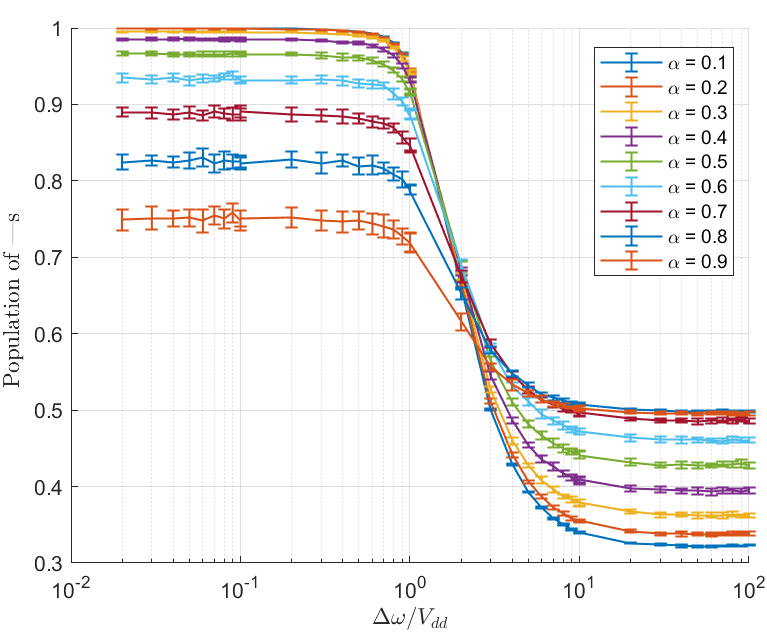}
\caption{Target state population $P_s$ versus normalized bandwidth under white phase noise for $\alpha = 0.1$--$0.9$ ($\alpha \equiv \epsilon_\phi$). Phase fluctuations produce significant population degradation even in the blockade regime. The critical threshold is $\epsilon_\phi \approx 0.01$, beyond which population drops below 90\%. Phase noise sensitivity exceeds amplitude noise sensitivity by over an order of magnitude.}
\label{fig:phase_noise}
\end{figure}

In stark contrast, phase fluctuations produce rapid fidelity degradation (Fig.~\ref{fig:phase_noise}). The critical threshold is $\alpha \equiv \epsilon_\phi \approx 0.01$ (1\% phase noise), beyond which population drops precipitously below~90\%.

Phase noise is fundamentally different because it accumulates coherently. The instantaneous phase error $\delta\phi(t) = \epsilon_\phi\eta_\phi(t)$ contributes to the total acquired phase $\Phi_{\mathrm{total}} = \omega_0 T + \delta\phi(T) - \delta\phi(0)$. White noise, with rapid fluctuations, produces large phase excursions that destroy coherence between pathways.

Here we find a significant spectral dependence. Pink noise produces consistently higher fidelity than white noise at matched amplitudes, because its $1/f$ spectral weighting concentrates power at low frequencies, producing slowly varying phase drifts that act as small detuning errors rather than incoherent dephasing. The OU noise results interpolate between white and pink depending on the correlation time $\tau_c$ relative to the pulse duration~$T$: when $\tau_c \ll T$, OU noise behaves like white noise; when $\tau_c \gtrsim T$, it approaches the pink noise limit.

\subsection{Model Comparison}
\label{subsec:model_comparison}

The three models (3LN, 3LA, 2LA) show consistent qualitative trends but quantitative differences, particularly at high noise levels. The full three-level model (3LN) generally predicts slightly lower fidelity, as the intermediate state $|e\rangle$ provides an additional pathway for population loss. The two-level approximation (2LA) provides qualitative insight for rapid parameter exploration, while the 3LN model is required for quantitative predictions. The Magnus-expansion-based analytical model (3LA) is accurate for narrow-band pulses but breaks down when higher-order terms become significant at large bandwidths.

\section{Optimal Control Under Noise}
\label{sec:qoct}

\subsection{Optimization Protocol}
\label{subsec:optimization}

We apply the D-MORPH algorithm described in Sec.~\ref{subsec:qoct} to design optimized pulses starting from Gaussian seed pulses at each duration ($\tau = 100$, 250, and 400~fs). The optimization is performed under ideal (noise-free) conditions, maximizing the Bell-state fidelity $F = |\langle s|\psi(t_f)\rangle|^2$ while simultaneously enforcing all three equality constraints. The step size~$ds$ is adaptively reduced when the fidelity decreases, ensuring monotonic convergence. We then evaluate the robustness of the resulting optimized pulses by propagating the system under noise in a post-hoc Monte Carlo analysis. This approach separates the deterministic optimization from the stochastic robustness assessment.

\subsection{Optimized Pulse Structure}
\label{subsec:pulse_structure}

\begin{figure}[htbp]
\centering
\begin{subfigure}[t]{\columnwidth}
    \centering
    \includegraphics[width=\linewidth]{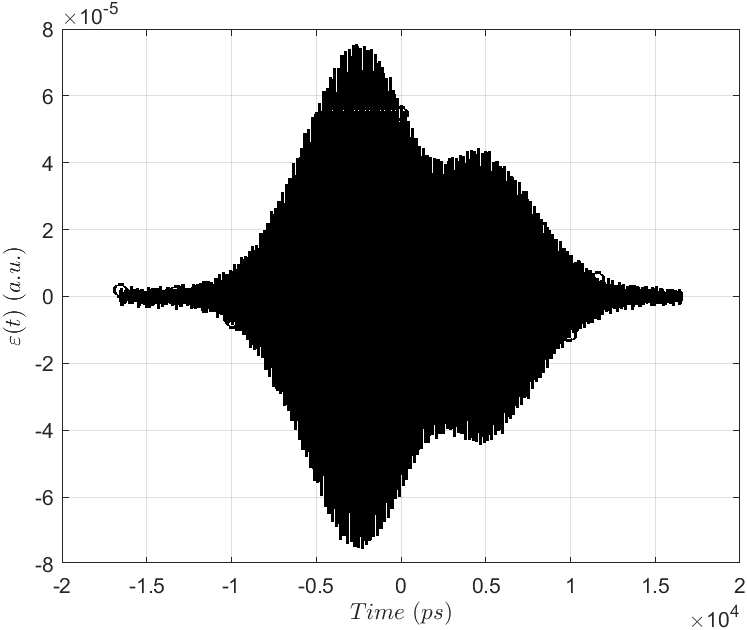}
    \caption{}
    \label{fig:optimal_pulse_time}
\end{subfigure}
\vspace{0.3cm}
\begin{subfigure}[t]{\columnwidth}
    \centering
    \includegraphics[width=\linewidth]{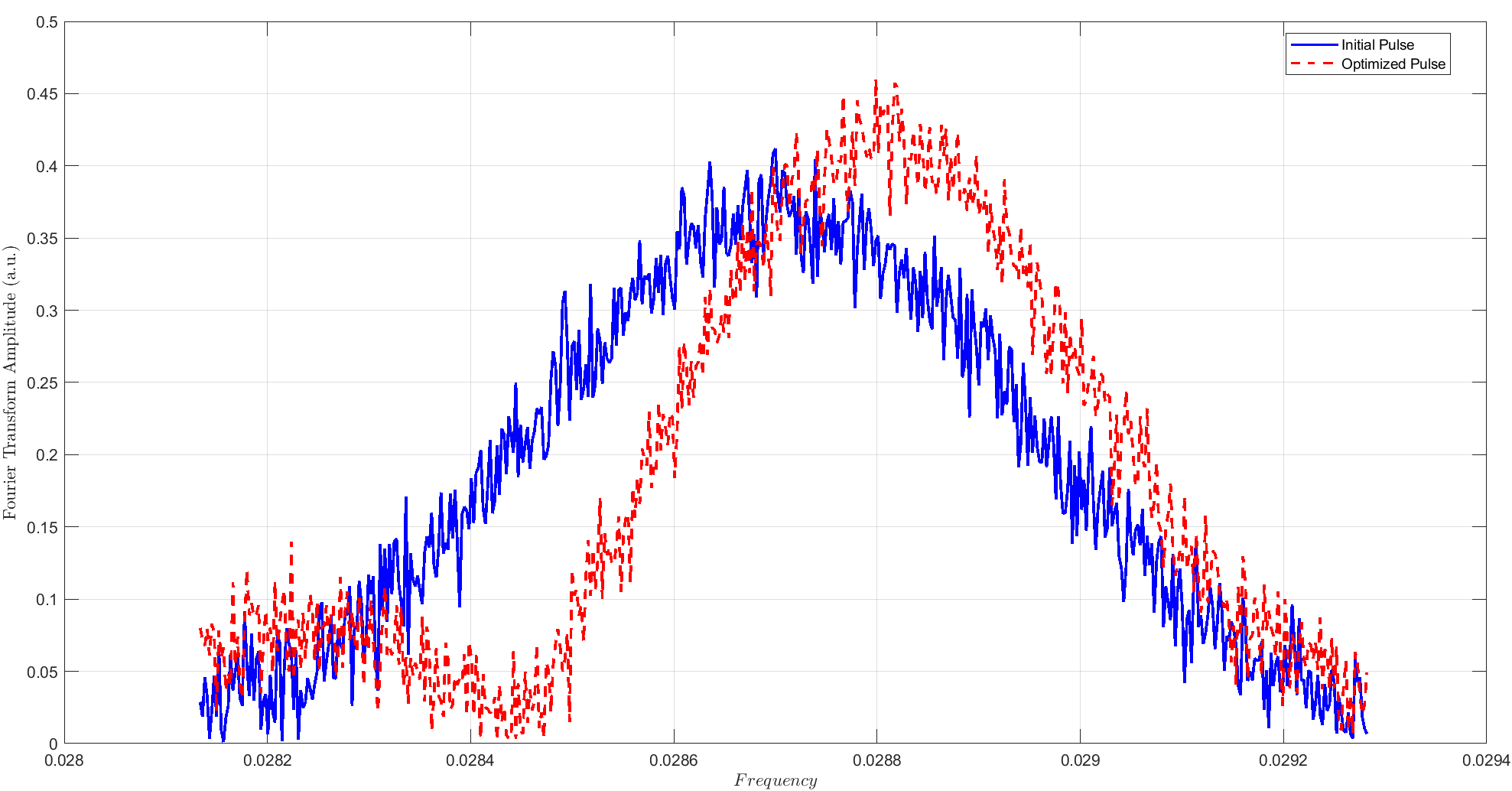}
    \caption{}
    \label{fig:optimal_pulse_spectrum}
\end{subfigure}
\caption{Optimized pulse structure from D-MORPH optimization. (a)~Time-domain electric field showing the characteristic double-pulse structure with two distinct sub-pulses separated by $\tau \approx \pi/\Vdd$. (b)~Frequency-domain spectra comparing initial Gaussian pulse (blue) and optimized pulse (red). The optimized spectrum is modified near the $\omega_{es}$ transition frequency, suppressing leakage to the doubly-excited state $|e\rangle$, while the spectral amplitude at $\omega_{sg}$ is preserved by the pulse-area constraint.}
\label{fig:optimal_pulse}
\end{figure}

The D-MORPH optimization converges to a characteristic double-pulse structure (Fig.~\ref{fig:optimal_pulse}). This structure emerges naturally from the algorithm; the temporal separation between the two sub-pulses is approximately $\pi/\Vdd$, corresponding to one half-cycle of the Rydberg interaction-induced oscillation. The optimized pulses achieve noise-free fidelities of $P_s \approx 0.76$ for $\tau = 100$~fs ($\Delta\omega \approx 4.0\,\Vdd$), $P_s > 0.99$ for $\tau = 250$~fs, and $P_s > 0.999$ for $\tau = 400$~fs, consistent with the published results of Ref.~\cite{Guo2019PRA}.

The spectral analysis reveals that the optimization achieves high fidelity by reducing the pulse area $\theta_{es}(t_f)$ while preserving $\theta_{sg}(t_f) = \pi/2$ via the equality constraint. Additionally, the higher-order terms in the Magnus expansion are modulated to further suppress population transfer to $|e\rangle$, as detailed in Ref.~\cite{Guo2019PRA}.

\subsection{Performance Under Noise}
\label{subsec:performance}

\begin{figure*}[htbp]
\centering
\begin{subfigure}[t]{0.32\textwidth}
    \centering
    \includegraphics[width=\linewidth]{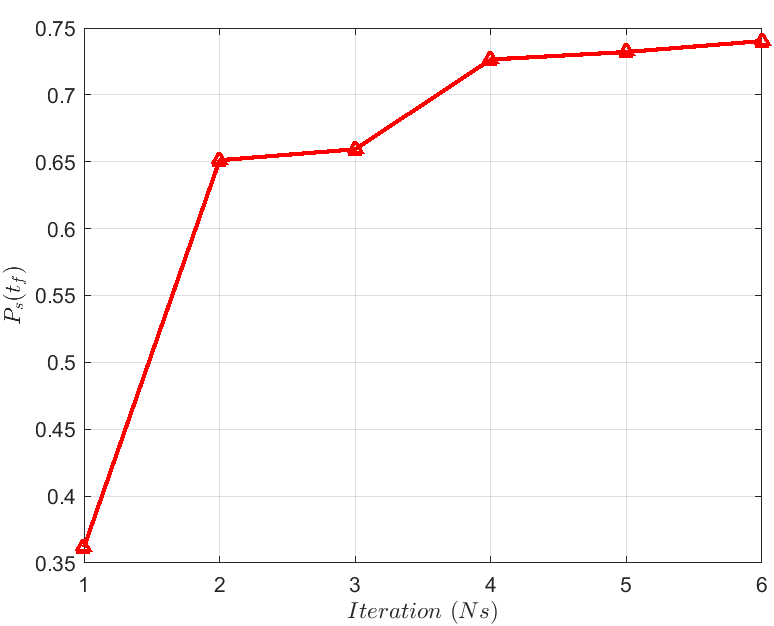}
    \caption{}
    \label{fig:fidelity_iteration}
\end{subfigure}
\hfill
\begin{subfigure}[t]{0.32\textwidth}
    \centering
    \includegraphics[width=\linewidth]{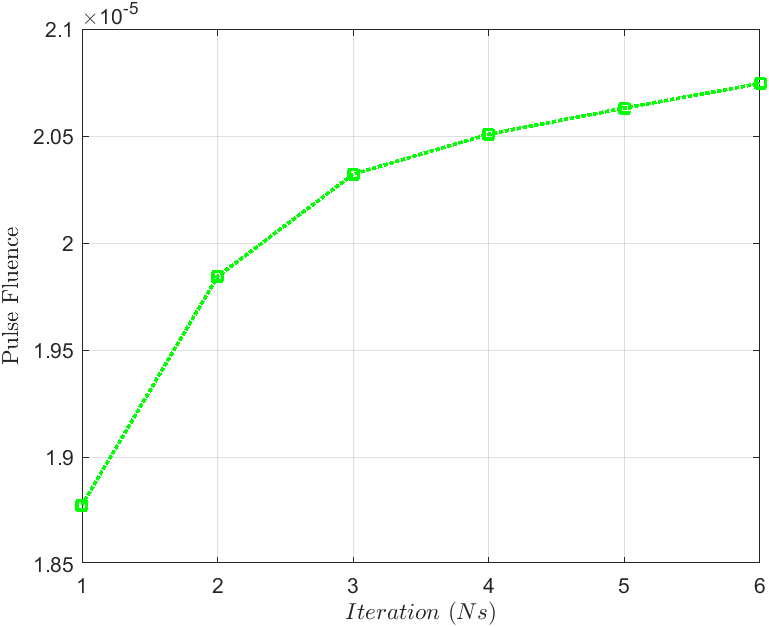}
    \caption{}
    \label{fig:fluence_iteration}
\end{subfigure}
\hfill
\begin{subfigure}[t]{0.32\textwidth}
    \centering
    \includegraphics[width=\linewidth]{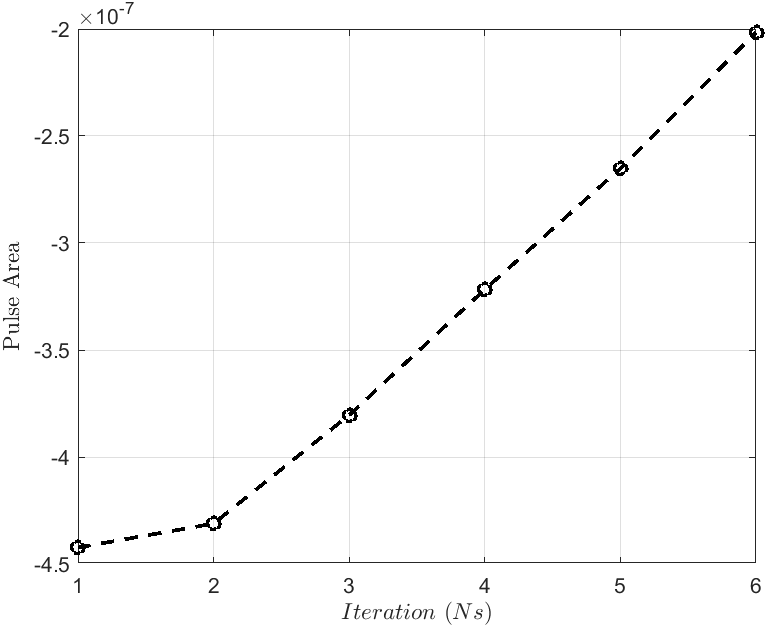}
    \caption{}
    \label{fig:area_iteration}
\end{subfigure}
\caption{D-MORPH optimization convergence ($\tau = 250$~fs). (a)~Target state population $P_s(t_f)$ versus iteration number, showing monotonic improvement to $P_s > 0.99$. (b)~Pulse fluence remains constant throughout optimization, confirming the energy-conservation constraint. (c)~Spectral pulse area: $\theta_{sg}$ is maintained at $\pi/2$ by the third constraint, while $\theta_{es}$ is systematically reduced, suppressing the unwanted $|s\rangle \to |e\rangle$ transition.}
\label{fig:optimal_performance}
\end{figure*}

Figure~\ref{fig:optimal_performance} shows the convergence of the D-MORPH optimization for the $\tau = 250$~fs case, demonstrating monotonic fidelity improvement, constant fluence (energy conservation), and preservation of the spectral pulse-area constraint throughout the iterations.

When the optimized pulses are evaluated under noise via Monte Carlo simulations, the results show clear advantages over simple Gaussian pulses. At low noise levels ($\epsilon_A \lesssim 10^{-3}$), both pulse types achieve near-unity fidelity. At moderate noise ($\epsilon_A \sim 0.01$), the optimized pulse maintains $F > 0.97$, while the Gaussian degrades more significantly. The optimized pulse also exhibits reduced fidelity variance, which is important for quantum error correction protocols requiring predictable gate performance.

\subsection{Breakdown Threshold}
\label{subsec:breakdown}

The optimized pulses exhibit a distinct breakdown threshold near $\epsilon_A \approx 0.01$ (1\% amplitude noise). Below this level, fidelity remains high; above it, performance degrades rapidly. This threshold has a clear physical interpretation: when noise-induced fluctuations in the Rabi frequency become comparable to the inverse pulse duration ($\epsilon_A \Omega_{\mathrm{peak}} \sim 1/T$), the coherent pulse area can no longer be reliably controlled, giving
\begin{equation}
\epsilon_A^{\mathrm{crit}} \sim \frac{1}{\Omega_{\mathrm{peak}} T}.
\label{eq:threshold}
\end{equation}
For our parameters, this predicts $\epsilon_A^{\mathrm{crit}} \sim 0.01$, consistent with the simulations.

\section{Discussion}
\label{sec:discussion}

\subsection{Comparison with Recent Experiments}

Our theoretical predictions can be contextualized against the rapid experimental progress in Rydberg quantum computing. Recent demonstrations have achieved two-qubit gate fidelities of 99.5\% using microsecond-timescale pulses~\cite{Evered2023Nature,Jandura2022}, and these experiments identify laser phase noise as a dominant error source~\cite{Levine2018,Jiang2023PRA}, consistent with our finding that phase noise sensitivity exceeds amplitude noise sensitivity by over an order of magnitude. The noise amplitudes achievable in current experiments ($\epsilon_A \lesssim 10^{-3}$, $\epsilon_\phi \lesssim 10^{-4}$) are well within the regime where our optimized pulses maintain high fidelity.

\subsection{Implications for Quantum Error Correction}

Fault-tolerant quantum computing requires two-qubit gate fidelities exceeding the surface code threshold ($\sim$99\%)~\cite{Fowler2012,Bluvstein2024Nature}. Our results indicate that amplitude noise at the 0.1\% level permits threshold-exceeding fidelities even with simple Gaussian pulses, that phase noise must be suppressed below $\sim$1\% to maintain threshold fidelities, and that optimized pulses provide meaningful fidelity improvements when phase noise is controlled. The variance reduction provided by optimal control is additionally valuable, as it reduces the occurrence of rare, high-error events that can trigger logical failures.

\subsection{Spectral Structure and Filter Functions}

Our finding that pink noise produces less fidelity degradation than white noise connects to the filter function formalism for quantum control~\cite{Oda2023,Khodjasteh2005}. The gate acts as a frequency-dependent filter on the noise, with the filter function $\mathcal{F}(\omega)$ determining how much noise power at frequency $\omega$ contributes to infidelity: $1 - F \propto \int d\omega\, S(\omega)|\mathcal{F}(\omega)|^2$. For ultrafast gates, $\mathcal{F}(\omega)$ is broad but suppressed at low frequencies due to the short gate time, naturally filtering out low-frequency-dominated pink noise more effectively than white noise.

\subsection{Extension to Multi-Qubit Systems}

The principles demonstrated here extend to multi-qubit Rydberg arrays. In parallel gate operations~\cite{Evered2023Nature,Graham2022}, correlated noise across multiple qubits becomes important~\cite{Lin2024}. Our spectral analysis suggests that engineering noise correlations to be predominantly at low frequencies would minimize crosstalk-induced errors.

\section{Conclusion}
\label{sec:conclusion}

We have presented a systematic theoretical analysis of ultrafast entanglement generation between Rydberg-blockaded atoms under realistic laser noise conditions. Entanglement fidelity is remarkably robust to amplitude fluctuations, maintaining above 90\% fidelity even at 30\% noise levels, owing to pulse-area averaging effects. In contrast, phase fluctuations are the critical limiting factor, with fidelity degrading rapidly beyond $\sim$1\% noise amplitude. The spectral structure of the noise matters: pink ($1/f$) noise consistently produces less fidelity degradation than white noise, suggesting that engineering noise spectra toward low-frequency dominance is beneficial.

D-MORPH optimization under three simultaneous equality constraints produces double-pulse structures that achieve $>$99\% fidelity in the ideal case and maintain superior noise tolerance compared to Gaussian pulses. A clear breakdown threshold near 1\% amplitude noise marks the limit of coherent control, establishing practical benchmarks for experimental implementation. These results provide quantitative guidance for the development of ultrafast neutral-atom quantum processors. Future work should address extension to multi-qubit gates, effects of atomic motion, and connections to open-system dynamics.

\begin{acknowledgments}
The author is grateful to Chuan-Cun Shu for valuable guidance, supervision, and helpful discussions throughout this work. This research was supported in part by the National Natural Science Foundation of China (NSFC). Computational resources were provided by Central South University.
\end{acknowledgments}

\bibliography{references}

\appendix

\section{Noise Generation Algorithm}
\label{app:noise}

Colored noise realizations are generated through spectral shaping. Starting from white noise $w_n \sim \mathcal{N}(0,1)$ in the time domain, we compute its FFT $\tilde{w}_k$, apply a spectral filter $\tilde{\eta}_k = \tilde{w}_k \sqrt{S(\omega_k)}$ for the target power spectral density $S(\omega)$, transform back via inverse FFT, and normalize to unit variance. For pink noise, the Voss--McCartney algorithm is used to avoid spectral leakage at low frequencies.

\section{D-MORPH Implementation Details}
\label{app:dmorph}

The D-MORPH update equation~(\ref{eq:dmorph}) is integrated iteratively with an adaptive step size $ds$. At each iteration, the time-dependent Schr\"{o}dinger equation is solved via matrix exponentiation $U(t_{j+1},t_j) = \exp(-i\,H_d(t_j)\,\Delta t)$ to propagate the wavefunction forward and the costate backward, from which the gradient $\delta F/\delta\mathcal{E}$ is computed via Eq.~(\ref{eq:gradient}). The Gram matrix~$\Lambda$ is assembled and inverted to obtain the constrained update direction. The step size is reduced by a factor of 10 if the fidelity decreases, ensuring monotonic convergence. Optimization terminates when the target fidelity is reached or a maximum number of iterations is exceeded.

\section{Model Comparison Under Optimized Pulses}
\label{app:model_comparison}

\begin{figure}[htbp]
\centering
\includegraphics[width=\columnwidth]{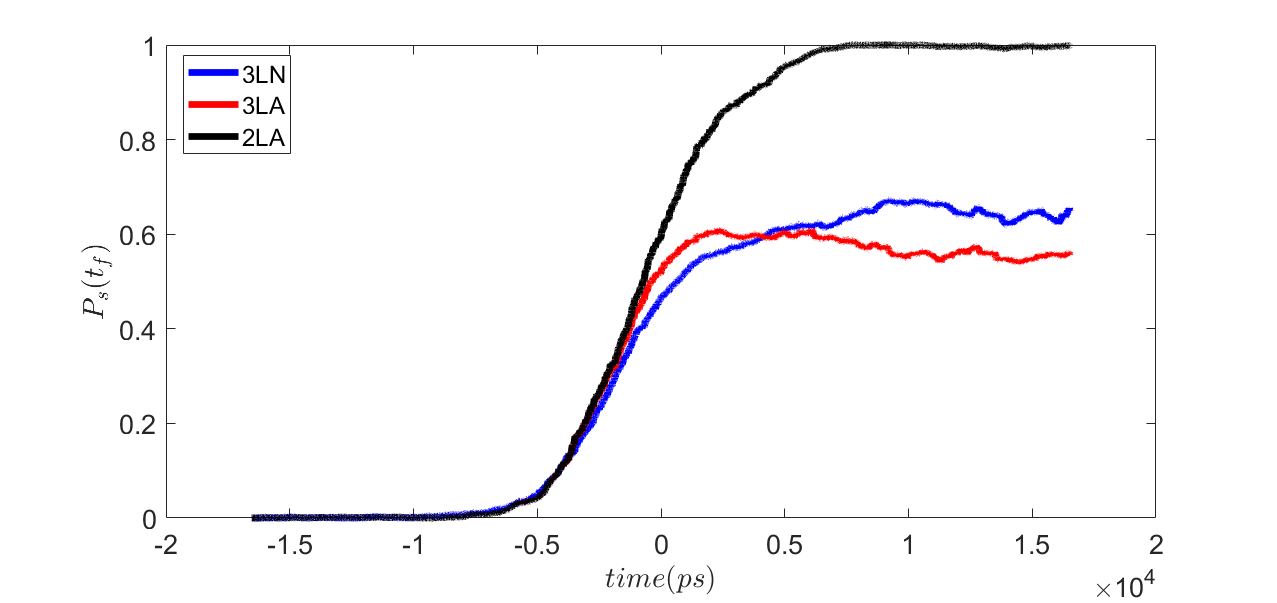}
\caption{Time-dependent population $P_s(t)$ for the three computational models (3LN: blue solid; 3LA: red dashed; 2LA: black dotted) under moderate white amplitude noise applied to the optimized pulse ($\tau = 100$~fs). The 2LA model predicts near-unity transfer, while 3LA and 3LN show progressively more realistic dynamics reflecting intermediate-state leakage.}
\label{fig:model_comparison_moderate}
\end{figure}

\begin{figure}[htbp]
\centering
\includegraphics[width=\columnwidth]{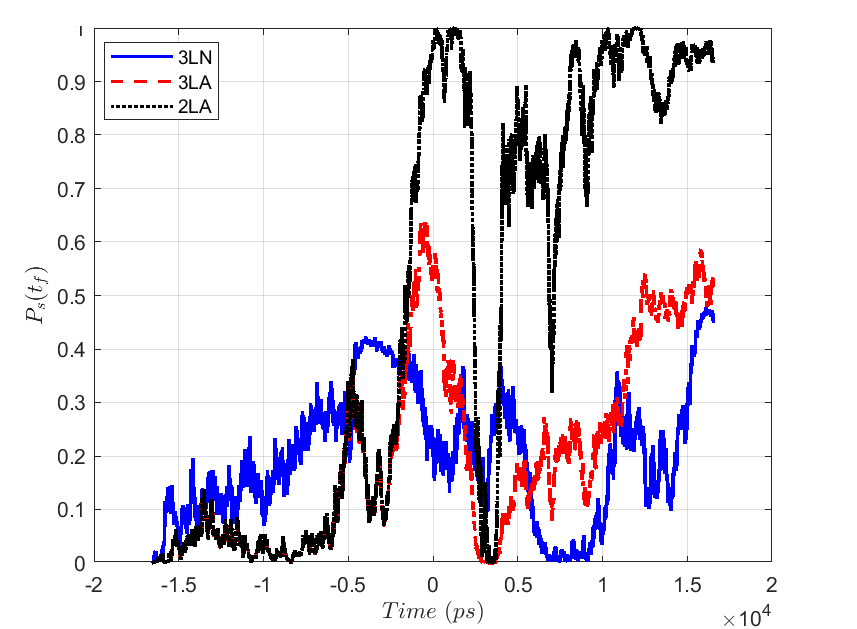}
\caption{Same as Fig.~\ref{fig:model_comparison_moderate} but with higher noise amplitude, illustrating the breakdown regime. The 2LA model shows unphysical oscillations reflecting its inadequacy at high noise, while 3LA and 3LN reveal substantial population loss, establishing the practical limits of optimized pulse performance.}
\label{fig:model_comparison_high}
\end{figure}

Figures~\ref{fig:model_comparison_moderate} and~\ref{fig:model_comparison_high} illustrate the population dynamics in $|s\rangle$ for all three models under white amplitude noise applied to D-MORPH optimized pulses at $\tau = 100$~fs, where the three model descriptions show the largest discrepancies (the high-fidelity results in the main text correspond to $\tau = 250$--400~fs).

Under moderate noise (Fig.~\ref{fig:model_comparison_moderate}), the 2LA model predicts near-ideal transfer ($P_s \approx 0.99$) since it neglects $|e\rangle$; the 3LA model shows intermediate behavior with some population loss; and the 3LN model captures the full dynamics including non-adiabatic corrections. At higher noise levels (Fig.~\ref{fig:model_comparison_high}), the 2LA model becomes unreliable, exhibiting unphysical oscillations, while the 3LN model reveals the true extent of fidelity degradation. These results confirm that the full 3LN treatment is essential for quantitative predictions under realistic noise conditions, and that the breakdown threshold identified in Sec.~\ref{subsec:breakdown} corresponds to the regime where coherent control is lost across all model descriptions.

\end{document}